\documentclass[english,aps,preprint]{revtex4-1}
\usepackage[T1]{fontenc}
\usepackage[latin9]{inputenc}
\usepackage{babel}
\begin{document}

\title{Comments on turbulence theory by Qian and by Edwards and McComb}

\author{R. V. R. Pandya}

\email{rvrptur(AT)yahoo.com}

\affiliation{Department of Mechanical Engineering, University of Puerto Rico at
Mayaguez, PR 00682, USA}
\begin{abstract}
We reexamine Liouville equation based turbulence theories proposed
by Qian {[}Phys. Fluids \textbf{26}, 2098 (1983){]} and Edwards and
McComb {[}J. Phys. A: Math. Gen. \textbf{2}, 157 (1969){]}, which
are compatible with Kolmogorov spectrum. These theories obtained identical
equation for spectral density $q(k)$ and different results for damping
coefficient. Qian proposed variational approach and Edwards and McComb
proposed maximal entropy principle to obtain equation for the damping
coefficient. We show that assumptions used in these theories to obtain
damping coefficient correspond to unphysical conditions. 
\end{abstract}
\maketitle

\section{Introduction}

Edwards \cite{Edwards64} proposed turbulence theory based on Liouville
equation for joint probability distribution function of Fourier modes
$u_{\alpha}({\bf k},t)$ of velocity field governed by forced Navier-Stokes
equation. Following Edwards, a few more turbulence theories \cite{EM69,Herring65,Qian83}
were proposed to solve Liouville equation and were reviewed by Leslie
\cite{Leslie73} and McComb \cite{McComb90}. Edwards theory seeks
Fokker-Planck model representation for Liouville equation and obtains
closed set of equations for spectral density $q(k)$ and damping coefficient
(total viscosity) $\omega(k)$ for stationary, isotropic turbulence.
The theory failed to be consistent with Kolmogorov spectrum \cite{Kolmogorov41,Batchelor59}
and the failure was attributed to equation for $\omega(k)$ \cite{McComb90}.
As a modification to Edwards' theory, Edwards and McComb \cite{EM69}
proposed principle of maximal entropy to derive an equation for $\omega(k)$
compatible with Kolmogorov spectrum. Within the Fokker-Planck framework
of Edwards, Qian \cite{Qian83} proposed variational approach and
obtained different equation for damping coefficient consistent with
Kolmogorov spectrum. Instead of using $u_{\alpha}({\bf k},t)$, Qian
used real dynamical modal variables $X_{i}$ and their governing equations
which were suggested by Kraichnan \cite{Kraichnan58} and later utilized
by Herring \cite{Herring65} for his self-consistent turbulence theory.
Herring's theory also uses Liouville equation and obtains equation
for $q(k)$ identical to equation obtained by Edwards. In this paper,
we reexamine Kolmogorov spectrum compatible theories proposed by Qian
\cite{Qian83} and Edwards and McComb \cite{EM69} to obtain damping
coefficient. We show in the next two sections that assumptions made
in these theories correspond to unphysical conditions.

\section{Variational Approach by Qian}

For discussion purpose, hereafter we refer to Qian's variational approach
as Q83. We use (Q83; \#) to represent equation number (\#) in Q83
paper \cite{Qian83}. Qian based his theory on governing equation
for real dynamical modal variables $X_{i}$ for stationary, homogeneous,
isotropic turbulence, written as 

\begin{equation}
\frac{dX_{i}}{dt}=-(\nu_{i}-\nu_{i}^{'})X_{i}+\sum_{j,m}A_{ijm}X_{j}X_{m},\quad(\mbox{Q83};8)\label{eq:7-1}
\end{equation}
where $\nu_{i}^{'}X_{i}$ represents external driving force. Einstein
summation convention of repeated indices is not utilized in Eq. (\ref{eq:7-1})
and in this section. It should be noted that the real dynamical modal
variables and their equations were first suggested by Kraichnan within
the context of hydromagnetic turbulence \cite{Kraichnan58}. Qian's
theory \cite{Qian83} seeks Langevin model representation for isotropic
turbulence in the form 

\begin{equation}
\frac{d}{dt}X_{i}\simeq-\eta_{i}X_{i}+f_{i},\quad\eta_{i}=\zeta_{i}+(\nu-\nu_{i}^{'})\quad(\mbox{Q83;}12)\label{eq:11}
\end{equation}
by using

\begin{equation}
\sum_{j,m}A_{ijm}X_{j}X_{m}\cong-\zeta_{i}X_{i}+f_{i}\quad(\mbox{Q83};11)\label{eq:12}
\end{equation}
 where $-\zeta_{i}X_{i}$ is dynamical damping term and $f_{i}$ is
white noise type forcing term. Qian proposed variational approach
to obtain damping coefficient $\eta_{i}$. The approach yields equation
for $\eta_{i}$ by minimizing a function $I(\eta_{i})$, written as

\begin{equation}
I=\sum_{i}\left\langle \left(\sum_{j,m}A_{ijm}X_{j}X_{m}-(-\zeta_{i}X_{i})\right)^{2}\right\rangle \quad(\mbox{Q83};29)\label{eq:13}
\end{equation}
and using 
\begin{equation}
\frac{\partial I}{\partial\eta_{i}}=0.\quad(\mbox{Q83};28)\label{eq:38}
\end{equation}
Here $\left\langle \,\,\right\rangle $ represents ensemble average.
Qian considered variation in $I$ under constraint $\phi_{i}=constant$,
where $\phi_{i}$ is related to $\left\langle X_{i}^{2}\right\rangle $
by

\begin{equation}
\left\langle X_{i}^{2}\right\rangle =\phi_{i}\left(1-\frac{\nu_{i}-\nu_{i}^{'}}{\eta_{i}}\right).\quad(\mbox{Q83};25)\label{eq:15}
\end{equation}
Also, $\phi_{i}$ is proportional to spectral density $q(k)$ \cite{Qian83}.

We now show that for $I=I(\eta_{i})$ and under the constraint of
$\phi_{i}=constant$,

\begin{equation}
\frac{\partial I}{\partial\eta_{i}}\ne0
\end{equation}
within the framework of Langevin model considered by Qian. Consequently,
the use of Eq. (\ref{eq:38}) to obtain $\eta_{i}$ is in error. For
stationary turbulence, solution of Eq. (\ref{eq:11}) suggests

\begin{equation}
2\eta_{i}\left\langle X_{i}^{2}\right\rangle =F_{i}\label{eq:8}
\end{equation}
 in which correlation of white noise forcing term 
\begin{equation}
\left\langle f_{i}(t)f_{i}(t')\right\rangle =F_{i}\delta(t-t')\label{eq:18}
\end{equation}
is utilized. Here $\delta(t-t')$ is Dirac delta function. Using Eqs.
(\ref{eq:12}) , (\ref{eq:13}) and (\ref{eq:18}), function $I(\eta_{i})$
can be written as

\begin{equation}
I=\sum_{i}\left\langle \left(\sum_{j,m}A_{ijm}X_{j}X_{m}-(-\zeta_{i}X_{i})\right)^{2}\right\rangle =\sum_{i}\left\langle f_{i}f_{i}\right\rangle =\delta(0)\sum_{i}F_{i}.\label{eq:19}
\end{equation}
 Further, using Eqs (\ref{eq:15}), (\ref{eq:8}), (\ref{eq:19})
and for $\phi_{i}=constant$, we can write

\begin{equation}
\frac{\partial I}{\partial\eta_{i}}=\delta(0)\sum_{j}\frac{\partial F_{j}}{\partial\eta_{i}}=\delta(0)\sum_{j}\frac{\partial2\eta_{j}\left\langle X_{j}^{2}\right\rangle }{\partial\eta_{i}}=2\delta(0)\phi_{i}.\label{eq:20-1}
\end{equation}
This Eq. (\ref{eq:20-1}) suggests that for all $i$ 
\begin{equation}
\frac{\partial I}{\partial\eta i}\ne0\label{eq:21}
\end{equation}
as $\phi_{i}\ne0$. In view of this, use of Eq. (\ref{eq:38}), i.e.
$\frac{\partial I}{\partial\eta_{i}}=0$, in Q83 to obtain $\eta_{i}$
is in error and corresponds to unphysical condition $\phi_{i}=0,\:\forall i$
for stationary turbulence.

Now we suggest possible modification within the framework of Q83.
Consider a function $V$, written as
\begin{equation}
V=\sum_{j}\frac{\partial I}{\partial\eta_{j}}=\sum_{j}2\delta(0)\phi_{j},\label{eq:13-1}
\end{equation}
which satisfies an exact condition 
\begin{equation}
\frac{\partial V}{\partial\eta_{i}}=0.\label{eq:14}
\end{equation}
This condition along with Eqs. (\ref{eq:13}) can be used, instead
of Eq. (\ref{eq:38}), to obtain equation for $\eta_{i}$.

\section{Maximal entropy principle by Edwards and McComb}

For discussion purpose, hereafter we refer to Edwards and McComb's
theory as EM69. We use (M90; \#) to represent equation number (\#)
in McComb's book \cite{McComb90}. Edwards and McComb \cite{EM69}
considered stationary, homogeneous, isotropic, turbulence inside a
cubic box of side $L$. Their Liouville equation based theory uses
equation for Fourier modes $u_{\alpha}({\bf k},t)$ of the velocity
field $u_{\alpha}({\bf x},t)$ governed by forced Navier-Stokes equation,
written as

\begin{equation}
\left(\frac{\partial}{\partial t}+\nu k^{2}\right)u_{\alpha}({\bf k},t)=M_{\alpha\beta\gamma}({\bf k})\sum_{{\bf j}}u_{\beta}({\bf j},t)u_{\gamma}({\bf {\bf k-j}},t)+f_{\alpha}({\bf k},t),\quad(\mbox{M90};4.81)\label{eq:17}
\end{equation}
where $f_{\alpha}$ represents external driving force, ${\bf k}$
is wavevector and $k^{2}=|{\bf k}|^{2}$. The Einstein summation convention
for repeated Greek indices is utilized while writing Eq. (\ref{eq:17})
and in this section. Edwards and McComb \cite{EM69} theory seeks
Fokker-Planck model equation for Liouville equation. The model equation
contains two model parameters, namely $r(k)$ and $s(k)$, which account
for contribution of nonlinear term in Eq. (\ref{eq:17}). The dynamical
damping coefficient $r(k)$ is related to damping coefficient $\omega(k)$
by

\begin{equation}
\omega(k)=\nu k^{2}+r(k).\quad(\mbox{M90};6.80)
\end{equation}
The coefficient $s(k)$ accounts for correlation of white noise forcing
in the Langevin equation for Fokker-Planck equation. Within the framework
of Edwards and McComb and for stationary turbulence, $\omega(k)$
and $s(k)$ are related by 
\begin{equation}
2\omega(k)q(k)=d(k)\quad(\mbox{M90};6.84)\label{eq:20}
\end{equation}
where
\begin{equation}
d(k)=W(k)+s(k)\quad(\mbox{M90};6.79)
\end{equation}
and $W(k)$ accounts for correlation of forcing term $f_{\alpha}({\bf k},t)$.
The spectral density $q(k)$ is defined by 

\begin{equation}
\left(\frac{2\pi}{L}\right)^{3}\left\langle u_{\alpha}({\bf k})u_{\beta}(-{\bf k})\right\rangle =D_{\alpha\beta}({\bf k})q(k),\quad(\mbox{M90};6.85)
\end{equation}
where $D_{\alpha\beta}({\bf k})=\delta_{\alpha\beta}-\frac{k_{\alpha}k_{\beta}}{|{\bf k}|^{2}}$.
EM69 considered entropy function $S=S[q(k),\omega(k)]$ to obtain
equation for $\omega(k)$ by maximizing $S$, corresponding to the
condition

\begin{equation}
\frac{\delta S}{\delta\omega(k)}+\sum_{{\bf j}}\left[\frac{\delta S}{\delta q(j)}\right]\frac{\delta q(j)}{\delta\omega(k)}=0.\quad(\mbox{M90};7.88)\label{eq:18-1}
\end{equation}
 Here
\begin{equation}
\frac{\delta q(j)}{\delta\omega(k)}=-\,\frac{d(k)\delta(k-j)}{2\omega^{2}(k)}+\frac{1}{2\omega(j)}\frac{\delta d(j)}{\delta\omega(k)}.\quad(\mbox{M90};7.89)\label{eq:45}
\end{equation}
and $\delta(k-j)=1$ when $k=j$ otherwise $\delta(k-j)=0$. Edwards
and McComb realized the difficulty in obtaining the second term on
the right-hand side (rhs) of Eq. (\ref{eq:45}). After neglecting
the second term, an approximate equation 
\begin{equation}
\frac{\delta q(j)}{\delta\omega(k)}=-\frac{d(k)\delta(k-j)}{2\omega^{2}(k)}\label{eq:45-1}
\end{equation}
was used for further calculation in EM69 \cite{McComb90}. This Eq.
(\ref{eq:45-1}) suggests that $\frac{\delta q(j)}{\delta\omega(k)}=0,\,\forall j\ne k$.
As a consequence, EM69 used following approximate equation 
\begin{equation}
\frac{\delta S}{\delta\omega(k)}-\left[\frac{d(k)}{2\omega^{2}(k)}\right]\frac{\delta S}{\delta q(k)}=0\quad(\mbox{M90};7.90)\label{eq:47}
\end{equation}
to obtain equation for $\omega(k)$. 

We now show that the neglect of the second term on the rhs of Eq.
(\ref{eq:45}) corresponds to unphysical condition. Consequently,
the use of Eq. (\ref{eq:47}) to obtain $\omega(k)$ is in error.
Since EM69 is proposed for stationary turbulence, 
\begin{equation}
\left(\frac{2\pi}{L}\right)^{3}\sum_{{\bf j}}\frac{1}{2}\left\langle u_{\alpha}({\bf j})u_{\alpha}(-{\bf j})\right\rangle =\sum_{{\bf j}}q(j)=Constant.\label{eq:26}
\end{equation}
and from which we can write exact equation 
\begin{equation}
\sum_{{\bf j}}\frac{\delta q(j)}{\delta\omega(k)}=\frac{\delta q(k)}{\delta\omega(k)}+\sum_{{\bf j},{\bf \,j}\ne{\bf k}}\frac{\delta q(j)}{\delta\omega(k)}=0.\label{eq:27}
\end{equation}
Substituting approximate Eq. (\ref{eq:45-1}) of EM89 into Eq. (\ref{eq:27})
and using Eq. (\ref{eq:20}), we obtain 
\begin{equation}
\frac{d(k)}{2\omega^{2}(k)}=\frac{q(k)}{\omega(k)}=0
\end{equation}
 and which is not correct for all $k$. In view of this, approximation
used in EM89 to obtain $\omega(k)$ corresponds to unphysical condition
$q(k)=0$ and does not comply with conservation of energy Eq. (\ref{eq:26})
for stationary turbulence where $q(k)\ne0,\,\forall k$.

It should be noted that, within the framework of EM69, the unphysical
behavior can be avoided if 

\begin{equation}
\frac{\delta S}{\delta\omega(k)}=0\label{eq:25}
\end{equation}
along with $q(k)=constant,\,\forall k$ is used instead of Eq. (\ref{eq:18-1}).
This means that $S=S(\omega(k))$ and second term on the rhs of Eq.
(\ref{eq:18-1}) is equal to zero and is neglected. This kind of neglect
by Qian in Q83 for function $I(\eta_{i})$ was considered mathematically
incorrect by McComb \cite{McComb90}. In our view, Eq. (\ref{eq:25})
can be considered as valid equation which seeks to optimize Entropy
when energy of turbulence remains constant as $q(k)=constant,\,\forall k$.

\section{Concluding remarks}

Within the Eulerian framework, a very few renormalized perturbation
theories of turbulence are consistent with Kolmogorov spectrum \cite{Leslie73,McComb90}.
In this paper, we have reexamined two such theories proposed by Qian
\cite{Qian83} and Edwards and McComb \cite{EM69} and have revealed
hidden unphysical conditions in these theories. We have suggested
possible modifications, Eqs. (\ref{eq:13-1}), (\ref{eq:14}) and
(\ref{eq:25}), to these theories but have not explored their usefulness
in obtaining damping coefficient consistent with Kolmogorov spectrum.
This will be explored in a broader context of our future work on turbulence
theory development within the framework of Kraichnan's direct interaction
approximation \cite{Pandya14}. 

\bibliographystyle{plain}

\end{document}